\documentclass[a4paper]{mem}
\usepackage{natbib}
\usepackage{graphicx}
\usepackage{amsmath}
\usepackage[a4paper]{hyperref}
\idline{00}{0}
\catcode`\@=11
\def\gsim{\ifmmode{\mathrel{\mathpalette\@versim>}}
    \else{$\mathrel{\mathpalette\@versim>}$}\fi}
\def\lsim{\ifmmode{\mathrel{\mathpalette\@versim<}}
    \else{$\mathrel{\mathpalette\@versim<}$}\fi}
\def\@versim#1#2{\lower 2.9truept \vbox{\baselineskip 0pt \lineskip
    0.5truept \ialign{$\m@th#1\hfil##\hfil$\crcr#2\crcr\sim\crcr}}}
\catcode`\@=12
\def\cRe{\langle R\rangle _{\rm e}}
\def\Dt{\Delta t}
\def\Dtmax{\Delta t_{\rm max}}
\def\Dtmin{\Delta t_{\rm min}}

\def\rme{r_{\rm M}}

\def\sg0{\sigma_0}

\def\Td{T_{\rm dyn}}

\def\ku{k_1}

\def\kt{k_3}

\def\alphatol{\alpha_{\rm tol}}
\def\thetamin{\theta_{\rm min}}
\begin{document}
   \title{A parallel implementation of a new fast algorithm for N-body
   simulations
%\thanks{this is a place for a title footnote}
}

   \author{P. Londrillo \inst{1}, 
   C. Nipoti \inst{2},\\ 
          \and
           L. Ciotti \inst{2}%\fnmsep
%\thanks{this is a place for placing a footnote in the author field }
}

   \offprints{P. Londrillo} \mail{Osservatorio Astronomico di Bologna,
via Ranzani 1, 40127 Bologna }

   \institute{INAF -- Osservatorio Astronomico di Bologna, 
         via Ranzani 1, 40127 Bologna, Italy \email{londrillo@bo.astro.it}\\ 
              \and Dipartimento di Astronomia, Universit\`a di Bologna,
         via Ranzani 1, 40127 Bologna, Italy 
              \email{nipoti@bo.astro.it, ciotti@bo.astro.it} \\
                           }
   \abstract{

   A new, momentum preserving fast Poisson solver for N-body systems
   sharing effective $O(N)$ computational complexity, has been
   recently developed by Dehnen (2000, 2002). We have implemented the
   proposed algorithms in a Fortran-90 code, and parallelized it by a
   domain decomposition using the MPI routines. The code has been
   applied to intensive numerical investigations of galaxy mergers, in
   particular focusing on the possible origin of some of the observed
   scaling relations of elliptical galaxies.  We found that the
   Fundamental Plane is preserved by an equal mass merging hierarchy,
   while it is {\it not} in a scenario where galaxies grow by
   accretion of smaller stellar systems. In addition, both the
   Faber-Jackson and Kormendy relations are {\it not} reproduced by
   our simulations.

    }

   \authorrunning{P. Londrillo et al.}
   
   \titlerunning{N-body simulations of galaxy mergers} 

   \maketitle
%
%________________________________________________________________

\section{Introduction}

In the last few years, fast algorithms for computing N-body
interactions relying on the general multipole expansion techniques
have been developed.  These schemes, usually referred to as Fast
Multipole Methods (FMMs, see, e.g., Greengard \& Rokhlin 1987, 1997),
are designed to have $O(N)$ computational complexity. However, it has
been noted that only $O(N{\rm log}N)$ scaling is usually achieved in
numerical implementation (see, e.g., Capuzzo-Dolcetta \& Miocchi 1998,
hereafter CM98). The $O(N{\rm log}N)$ operation count also
characterizes the N-body codes based on tree algorithms, originally
proposed by Barnes \& Hut (1986, hereafter BH86). In addition, it has
been shown by direct numerical tests that, for given accuracy,
tree-codes are faster than the FMMs by a factor of $\sim 4$
(CM98). For this reason, most of the N-body codes currently used for
astrophysical applications are based on the classical BH86 tree
scheme, where the particle data are organized into a nested oct-tree
cell structure. Interactions between distant particles are
approximated by cell--particle interactions, and the distribution of
the particles on each cell is represented by a multipole expansion,
usually truncated to the quadrupole term.

Recently, a new scheme has been introduced (Dehnen 2000, 2002,
hereafter D02), which can be seen as an original combination of the
tree-based BH86 scheme and of the FMMs multipole expansion, truncated
at a fixed low order level (the proposed $p=3$ order results the
optimal one). This scheme, implemented by the Author in a C++ code
[named falcON, Force Algorithm with Complexity $O(N)$], results in a
significant improvement over the existing BH86 tree-codes, in terms of
uniform accuracy, linear momentum conservation, and CPU time
performances. Moreover, the numerical tests presented in the D02 paper
show that for an accuracy level $\epsilon \simeq 10^{-3}$, for the
first time an {\it effective} $O(N)$ scaling in operation count, as
predicted by the FMM theory, is in fact obtained.  For these reasons
two of us (P.L. and C.N.) have implemented this scheme in a new
Fortran-90 code (FVFPS, a Fortran Version of a Fast Poisson Solver),
and have then parallelized it by using the MPI procedures. A
completion of the code to handle gas-dynamics using Euler equations
and the AMR (Adaptive Mesh Refinement) techniques is under
development.  Here we briefly describe the main characteristics of
Dehnen's algorithm, and we present our FVFPS implementation and its
parallelization. The achieved accuracy level, the performances in
terms of operation count speed-up, the effective $O(N)$ scaling, as
well as the parallel scaling with the processors number, are then
documented in numerical tests. Finally, we describe some interesting
astrophysical applications of the code, in the context of the study of
galaxy merging.  Our version of the code is available upon request.

\section{Dehnen's scheme}

In the FMM approach, a system of particles is first organized in a
oct-tree structure of nested cells, providing a hierarchical grouping
of particles. For each pair of distant cells $(A,B)$, the interactions
between particles are then approximated by a Taylor expansion of the
two-point Green function around the {\it geometrical center} of each
cell by using spherical harmonics.  In these cell-cell interactions
the two-body symmetry of the exact Green function is preserved by the
multipole approximation.  In principle, a FMM-based scheme can achieve
a prescribed accuracy $\epsilon$ by increasing the order $p$ of the
Taylor polynomial, and/or by lowering the expansion parameter
$\eta\equiv r/R$ (where R is the distance between the two cell centers
and $r$ gives the typical size of a cell enclosing a particle group).
Even if a formal, asymptotic scaling $O(N)$ is expected, the proposed
algorithms in the original formulation (Greengard \& Rokhlin, 1997)
appear in fact to be less performing and of difficult
application. Therefore, a first variant has been introduced in CM98,
where the expansion order is kept fixed to $p=2$, and only the $\eta$
parameter controls the accuracy of the Taylor expansion.  This
approach takes advantage of the adaptivity of tree algorithms to
select the pairs of cells (even belonging to different levels of
refinements), which satisfy the opening criterium $r_{\rm crit}/R \leq
\theta$, where $\theta$ is a preassigned control parameter. To improve
the efficiency of the scheme, a new definition of the critical size
$r_{\rm crit}(A,B)$, has been also introduced. However, even in this
improved formulation, the FMM scheme results to be not competitive
with the classical BH86 tree-code.

Dehnen's scheme basically develops along similar lines, but with some
new ingredients, which in fact succeeded to improve the overall
efficiency in a substantial way:
\begin{itemize}

\item the Taylor polynomial expansion is performed in Cartesian
components, and the expansion centers are the {center of mass} of each
cell, as in the BH86 tree-code;

\item a fixed expansion order $p=3$ is chosen as optimal;

\item a new definition of the critical radius $r_{\rm crit}$ is
introduced for each cell, to optimize the adaptive selection of all
pairs of well-separated cells; 

\item a new, mass dependent control parameter $\theta=\theta(M)$ has
been introduced, that assures faster performances and a more uniform
error distribution.  The $\theta(M)$ function is given in implicit
form by
\begin{equation} 
{\theta^5 \over (1-\theta)^2} = {\thetamin^5 \over (1-\thetamin)^2}
{\left({ M \over M_{\rm tot} }\right)}^{- {1\over 3}}, 
\end{equation}
where $\thetamin$ (the minimum value of $\theta$, associated with the
total mass) enters as a new, preassigned control parameter, and where
$M$ is the total mass enclosed in a cell.

\end{itemize} 
The resulting force solver is then organized as follows: 

 (a) As in the BH86 code, an oct-tree structure of nested cells
containing particles is first constructed. A generic cell $A$, when
acting as a gravity ``source'' is characterized by the position of its
center of mass ${\bf X}_A$, by its critical size $r_A$, mass $M_A$,
and by the quadrupole tensor $Q_A$. For the same cell $A$, when acting
as a gravity ``sink'' a set of (twenty) Taylor coefficients $C_A$ are
used to store the potential and acceleration values at ${\bf X}_A$.

 (b) To compute these coefficients, a new procedure for the tree
exploration has been designed: starting from the root cell, all pairs
of cells $(A,B)$ satisfying the acceptance condition $(r_A+r_B)/R \leq
\theta$, are looked for in a sequential way.  When the condition is
satisfied, each cell of the selected pair accumulates the resulting
interaction contributions in $C_A(B)$ or $C_B(A)$. In case of nearby
cells (not satisfying the acceptance condition), the bigger cell of
the pair is splitted, and the new set of pairs are then
considered. However, in order to avoid visiting the tree too deeply,
some empirical conditions are adopted to truncate the sequence, by
computing directly (exact) two-body forces for particles in the
smallest cells.  The Author, by using a stack structure to order the
visited cell-pairs, has introduced a powerful and efficient new
algorithm able to perform this {\it interaction phase} in the force
computations.

 (c) Finally, once all the cell-cell interactions has been computed
and stored in the $C_A$ array, a final step is needed to evaluate the
gravitational potential and the accelerations of the particles in the
(generic) cell $A$. This {\it evaluation phase} can be performed by a
simple and fast $O(N)$ tree-traversal.

In our F-90 implementation of the scheme above, we followed rather
closely the algorithms as described in D02: at this stage of
development and testing, we have neither attempted to optimize our
implementation, nor to look for particular tricks to save
computational time.

To parallelize the force solver, we have adopted a straightforward
strategy, by first decomposing the physical domain into a set of
non-overlapping subdomains, each containing a similar number of
particles (a few per cent of tolerance is allowed).  The set of
particles on each subdomain, $S_k$, is then assigned to a specific
processing unit, $P_k$. Each processor builds its own tree [step (a)],
and computes the interaction phase for all the particles in its
subdomain [step (b)].  Steps (a) and (b) are executed in parallel. By
ordering the processors in a one-dimensional periodic chain, it is
then possible to exchange (particles and tree-nodes) data between
them, by a $``{\rm send}\_{\rm receive}''$ MPI routine, in such a way
that each processor can compute the interaction phase with the other
subdomains.  This phase is then performed by the following
computational steps:

$(i)$ The $P_k$ processor sends its source-data to $P_{k+s}$ and
receives from the $P_{k-s}$ a copy of its source-data, where
$s=1,2,...N_{\rm PE}/2$, and $N_{\rm PE}$ is the total number of
processors.

$(ii)$ At any given level $s$, each $P_k$ computes, again in
parallel, the interaction phase with particles and cells of
$P_{k-s}$. Mutuality of the interactions allows to evaluate, at the
same time, the results of these interactions for cells and particles
of processor $P_{k-s}$ to be stored as sink-data.  

$(iii)$ These accumulated sink-data (acceleration for particles and
Taylor's coefficients for cells) are sent back to the $P_{k-s}$
processor, while the corresponding sink-data of the $P_k$ processor
are received from processor $P_{k+s}$ and added to the current $P_{k}$
values.  In this way, it is evident that by repeating step $(ii)$
$N_{\rm PE}/2$ times and step $(iii)$ $N_{\rm PE}/2-1$ times, the
interactions of $P_k$ particles with all other particles in the domain
are recovered.

$(iv)$ The final evaluation phase [step (c)], being fully local, is
performed in parallel.

Besides the force solver, the N-body code has been completed by a
leap-frog time integrator scheme, with a (uniform) time step size
$\Delta t$ determined adaptively using a local stability condition
$\Delta t < 1/\sqrt{4\pi G\rho_{\rm max}}$, where $G$ is the
gravitational constant, and $\rho_{\rm max}$ the maximum of particle
density. For typical simulations involving galaxy mergers, this
adaptive (in time) step size keeps comparable to the value $\Td/100$,
where $\Td$ represents a macroscopic dynamical time scale, and assures
energy conservation with relative errors smaller than 0.1\%.

%                                                Two column figure
%----------------------------------------------------------- S_vib   
   \begin{figure*} \centering
   \resizebox{\hsize}{!}{\includegraphics[clip=true]{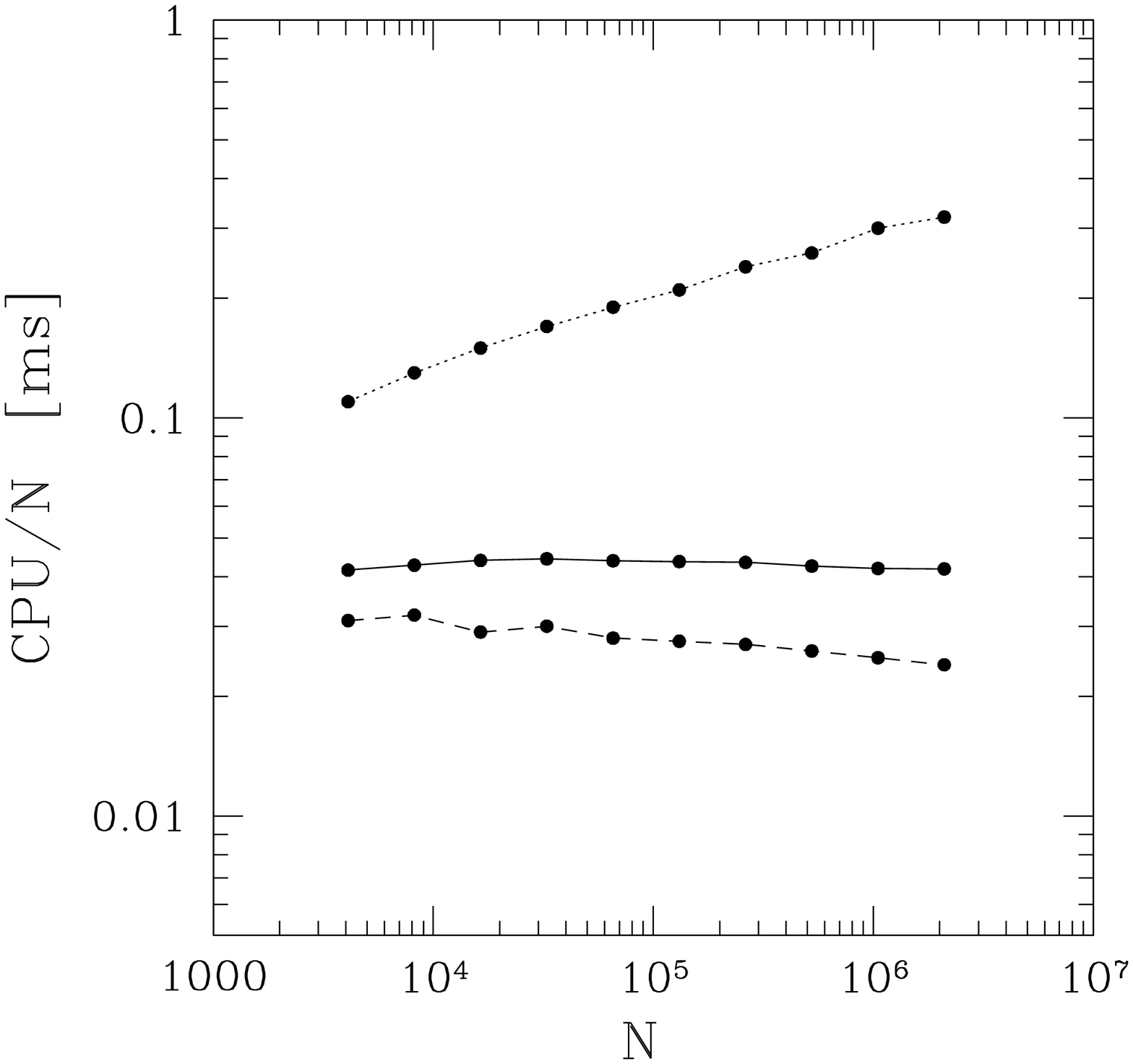}
   \includegraphics[clip=true]{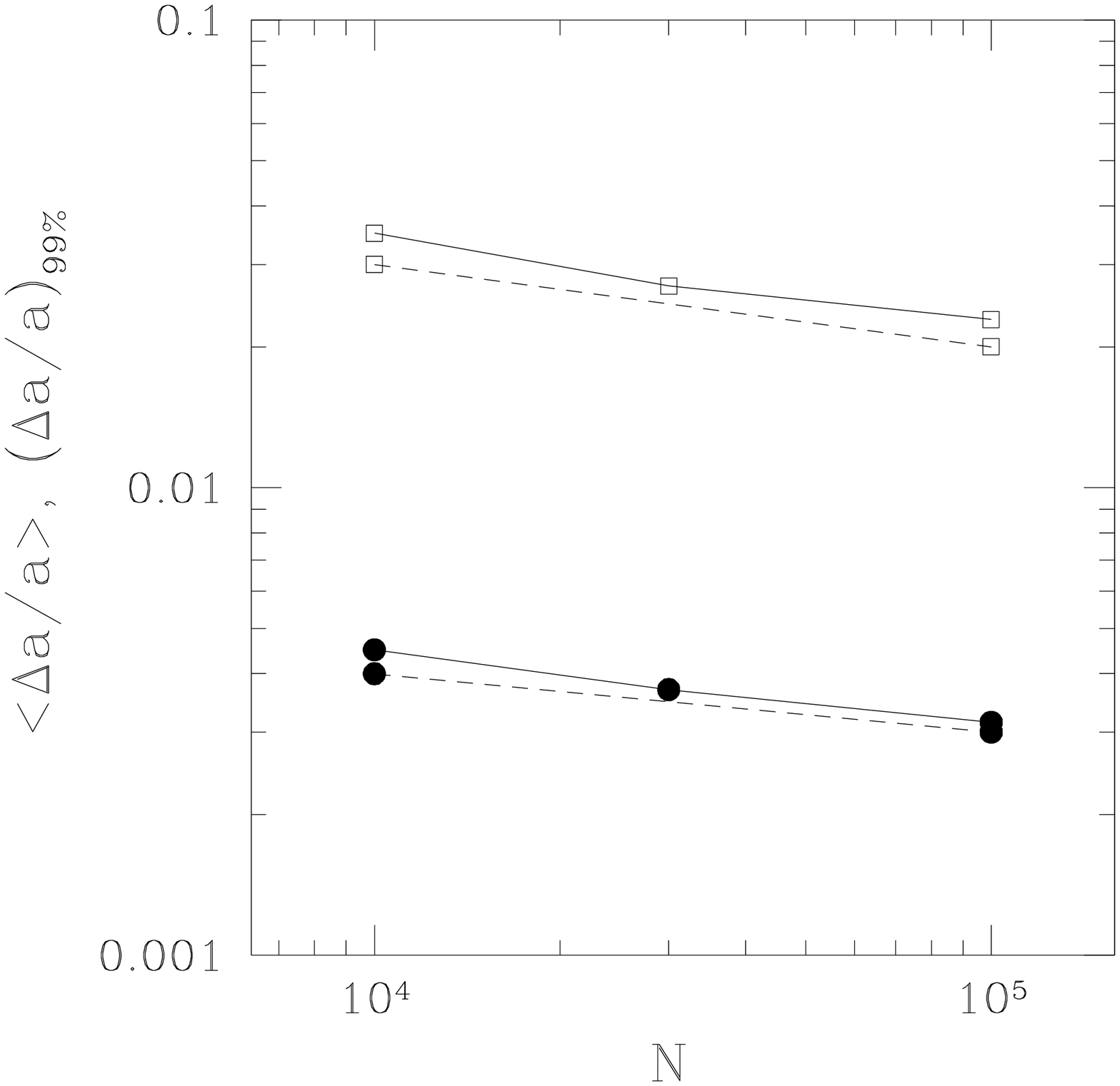}} \caption{{\it Left panel}:
   CPU time per particle as a function of total number of particles
   for BH86 tree-code (dotted line), falcON (dashed line; data from
   D02), and FVFPS (solid line).  For the BH86 code $\theta=0.8$,
   while in the last two cases $\theta=\theta(M)$, with
   $\thetamin=0.5$. {\it Right panel}: Mean error on the force
   calculation (filled circles) and the 99th percentile (empty squares),
   as a function of the number of particles in falcON (dashed line;
   data from D02), and in FVFPS (solid line).}  \label{fit1}
   \end{figure*}
%------------------------------------------------

\section{Performances of the FVFPS code}

We ran some test simulations in order to estimate the performances of
our FVFPS code, and to compare it both with falcON and with the BH86
tree-code.  We focus first on the analysis of the time efficiency in
force calculation, considering (one--processor) {\it scalar
simulations} of the N-body system representing a galaxy group, where
each galaxy is initialized by a Hernquist (1990, hereafter H90) galaxy
model. Values of the total number of particles $N$ up to $2 \times
10^6$ are considered. The opening parameter given in eq. (1) is used
by adopting $\thetamin=0.5$, while for the BH86 code we used
$\theta=0.8$: this choice results in comparable accuracies in the
force evaluation.  We ran our simulations on a Pentium III/1.133 GHz
PC.  For falcON and for the BH86 code we refer to the data published
by D02, who used a Pentium III/933 MHz PC.

In Fig. 1 (left panel) we plot the time efficiency of the codes as a
function of the number of particles; it is apparent from the diagram
that FVFPS (solid line), as well as falcON (dashed line), reaches an
effective $O(N)$ scaling, for $N \gsim 10^4$.  Our F-90 implementation
results to be slower than the C++ version by a factor of $\sim 1.5$:
we think that this discrepancy may be due to differences in the
details of the implementation, and also reflects a better intrinsic
efficiency of the C++ with respect to the F-90 for this kind of
algorithms (especially for routines where intensive memory accessing
is required). In the same diagram the data relative to the BH86
tree-code are also plotted (dotted line), and the $O(N{\rm log}N)$
scaling is apparent.  As already documented in D02, it is remarkable
how, for fixed $N$ and prescribed accuracy, the BH86 tree-code is
significantly slower than the new scheme: for example for
$N\simeq10^5$ the FVFPS code is faster by a factor of $\sim
8$. Therefore, even if the present F-90 implementation does no yet
appear to be optimal, it gives further support to the efficiency of
the D02 scheme, independently of the adopted programming language and
of the implementation details.

%______________________________________________________________
%\newpage
   \begin{figure*} \centering
   \resizebox{\hsize}{!}{\includegraphics[clip=true]{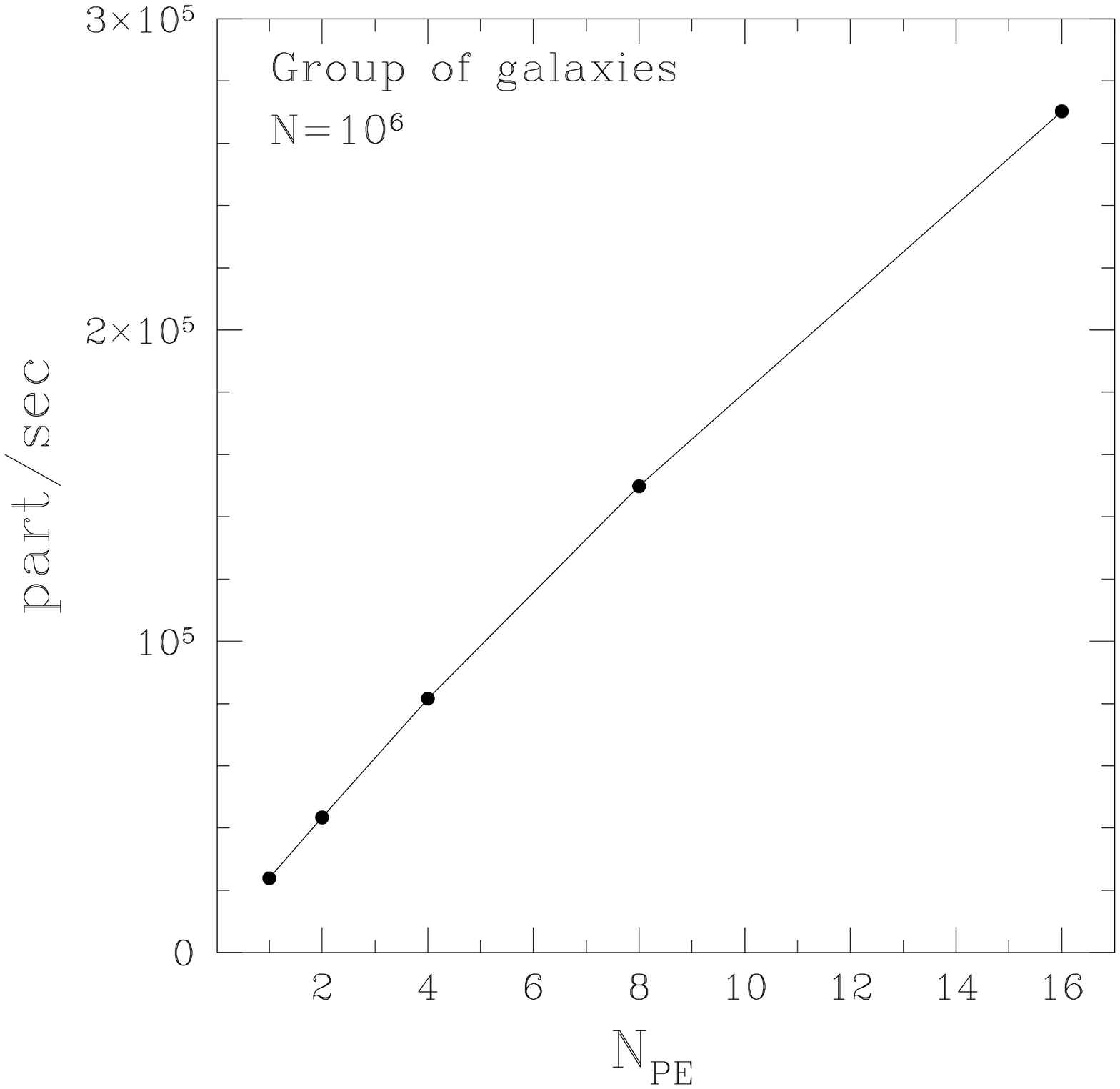}
   \includegraphics[clip=true]{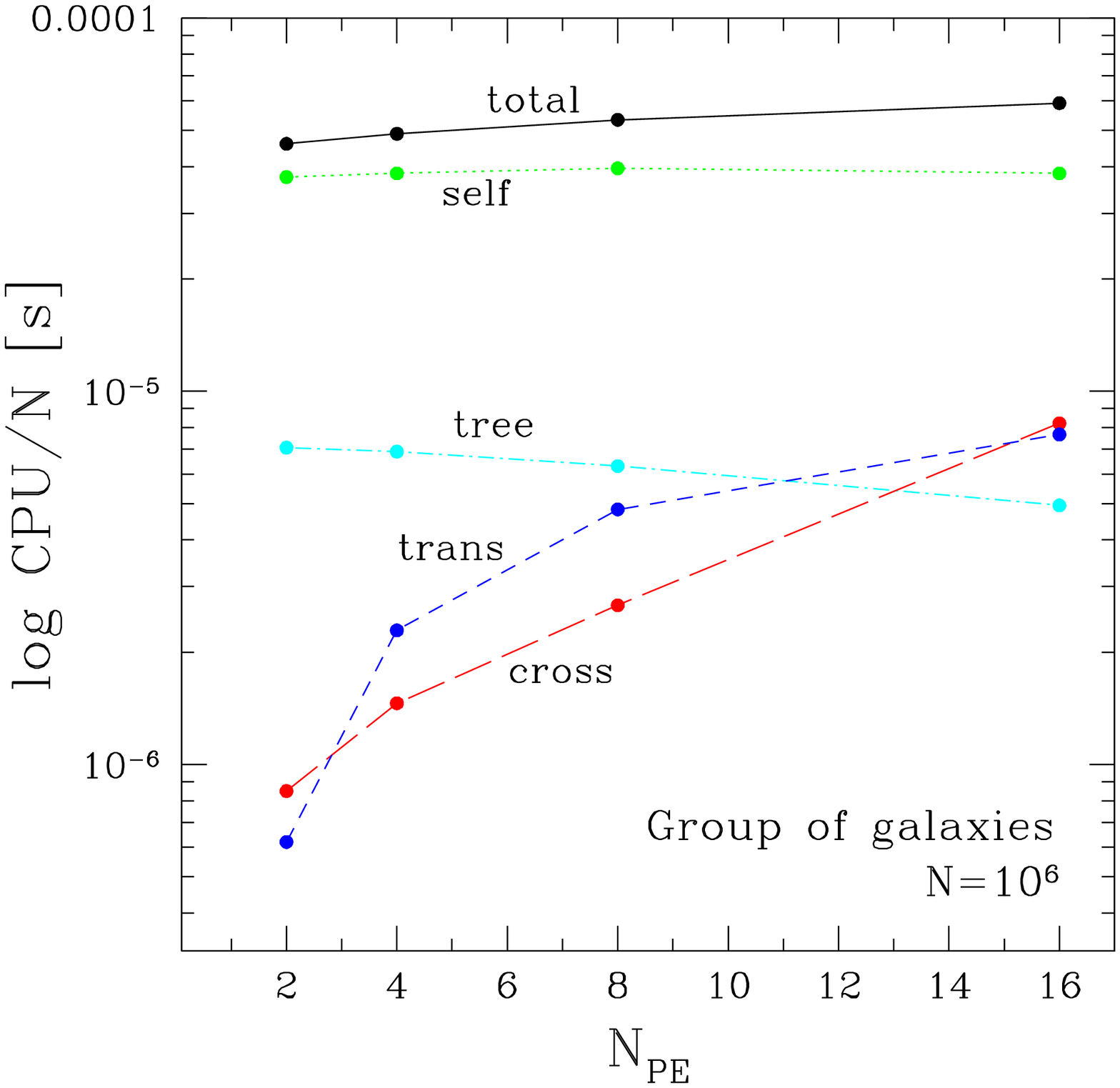}} \caption{{\it Left
   panel}: number of particle processed per second by the parallel
   version of the FVFPS code, as a function of the number of
   processors. The data refer to simulations of a group of galaxies
   with total number of particles $N=1048576$. {\it Right panel}: CPU
   time per particle as a function of the number of processors for the
   same simulations in left panel. The different contributions are
   plotted (see text for an explanation).} \label{fig2} \end{figure*}

%%%%%%%%%%%%%%%%%%%%%%%%%%%%%%%%%-------------------------------

The accuracy in force evaluation is documented on the right panel of 
Fig. 1, where we plot, as a function of $N$, the
mean relative error on the acceleration modulus (filled circles)
\begin{equation}
\left< {\Delta a \over a} \right> \equiv {1 \over N} 
\sum_{i=1}^N {|a^i_{\rm approx}-a^i_{\rm direct}| \over a^i_{\rm direct}},  
\end{equation}
and the corresponding 99th percentile $(\Delta a/a)_{99\%}$ (empty
squares), for FVFPS (solid lines) and, for comparison, in 
falcON (dashed lines). As usual, the reference
acceleration values $a^i_{\rm direct}$ are evaluated by direct summation of the
exact two-body Green function, while the approximated values are given
with $\theta=\theta(M)$, and $\thetamin=0.5$. In this case, we have considered
only simulations having a single H90 galaxy model.
From the diagram it is apparent that, for both implementations,  
$\langle \Delta a/a \rangle $ and $(\Delta a/a)_{99\%}$ do not exceed a few 
$10^{-3}$ and a few per cent, respectively, assuring accuracies usually 
considered sufficient for astrophysical applications (see, e.g., CM98).

As outlined in the previous Section, we parallelized our code by using
a domain decomposition with a load balancing essentially based only on
particles number on each subdomain. This load balancing criterium
seems in fact to work well, taking advantage of the intrinsic $O(N)$
scaling in the operation count of the basic scheme.  In order to check
the performance of the parallelized algorithm, we ran some simulations
on the IBM Linux cluster (Pentium III/1.133 GHz PCs) at the CINECA
center, in a range $N_{\rm PE} \leq 16$ in the number of processors.
In this case the initial conditions are represented by a group of 8
one--component H90 galaxy models with a total number of particles
$N=1048576$. As can be observed in Fig. 2 (left panel), this
parallelization is characterized by a rather well behaved linear
scaling of the CPU time per particle versus $N_{\rm PE}$.  This
scaling can be better evaluated by considering the right panel of
Fig. 2, where different contributions to the total CPU consumption are
plotted. As expected, up to $N_{\rm PE}=16$ the main contributions
come by the tree building (``tree'') and by the the force computation
in each sub-domain (``self''), corresponding to the step $(i)$ and
$(ii)$, respectively, of the parallel scheme outlined in the previous
Section. The contributions coming from the force evaluation among
particles belonging to different subdomains (``cross'') and the
cumulative time for all communication steps among processors
(``trans''), which increase with $N_{\rm PE}$, keep smaller in the
chosen range of parameters $(N,N_{\rm PE})$. Therefore, the $O(N)$
scaling of the base scheme, allows to estimate, for given $N$, the
optimal configuration of the $N_{\rm PE}$ that can be used.

\section{Simulations of galaxy merging}

As an astrophysical application of the described FVFPS code, we
consider here the study of the galaxy merging process. We checked the
reliability of the numerical simulations performed with FVFPS for this
kind of application (now time evolution is also considered), by
running a few merging events, between spherically symmetric stellar
systems, both with GADGET (Springel, Yoshida \& White 2001), and with
FVFPS. The results are in remarkable agreement: the remnants obtained
with the two codes are practically indistinguishable for
three-dimensional shape, circularized half--mass radius, circularized
density and velocity dispersion profiles. As an illustrative example,
in Fig. 3 we show the density profiles of the end--product of the
merger obtained from the head--on parabolic encounter of two H90
models, with a total number of particles $N=65536$ obtained with
GADGET (empty symbols), and with FVFPS (filled symbols); note how also
``minor'' details, as the slope change in the density profile at $r
\sim 10 \rme$, are nicely reproduced. In addition, in both cases the
change in total energy $\Delta E/E$ is smaller than $10^{-4}$ per
dynamical time, over $100\,\Td$. In this test, when using the FVFPS
code, we adopted softening parameter $\varepsilon=0.05$, initial time
step $\Dt=0.01$, and (constant) $\theta=0.65$; in the GADGET
simulation we used $\varepsilon=0.05$, $\alphatol=0.05$, $\Dtmin=0$,
$\Dtmax=0.03$ , $\alpha=0.02$ (see Springel et al. 2001 for details),
where the softening parameters and the time step parameters are in
units of the galaxy core radius and dynamical time, respectively.

%_____________________________________________________________
%                 A figure as large as the width of the column
%-------------------------------------------------------------
   \begin{figure} \centering
   \includegraphics[width=0.5\textwidth]{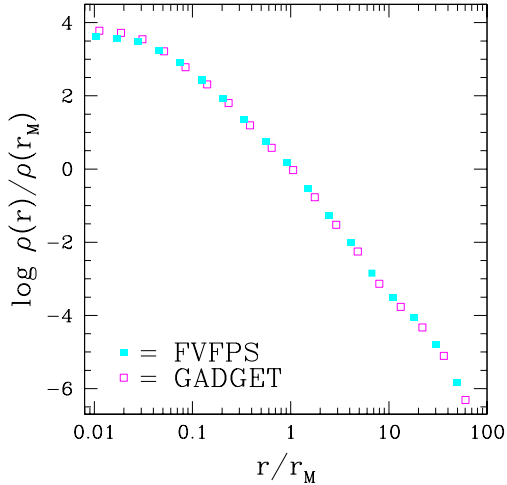}
   \caption{Angle--average density profile of the remnant of the
   merging of two identical H90 galaxy models with FVFPS (full
   symbols), and with GADGET (empty symbols). $\rme$ is the half--mass
   radius.} \label{FigVibStab} \end{figure}
%
%_____________________________________________________________
 
On the basis of the results of these tests, we used our N-body code to
investigate with numerical simulations the effects of hierarchies of
{\it dissipationless} galaxy mergers on the scaling relations of
elliptical galaxies. This study is described in details in Nipoti,
Londrillo \& Ciotti (2002, hereafter NLC02). Here we report, as an
example, the results relative to the effects of merging on the
Fundamental Plane of elliptical galaxies (hereafter FP; Djorgovski \&
Davis 1987, Dressler et al. 1987). We considered a hierarchy of 5
equal mass mergers and a hierarchy of 19 accretion events,
corresponding to an {\it effective} mass increase of a factor of $\sim
24$ and $\sim 16$, respectively: in the former case the successive
generations are produced by merging pairs of identical systems
obtained by duplicating the end--product of the previous step, while
in the latter case each end--product merges with a seed galaxy. Each
merging is produced in a head--on parabolic encounter. The seed
galaxies are spherically symmetric one--component H90 galaxy models,
with number of particles $N=16384$ each. As a consequence of the
merging hierarchy, the number of particles in the simulations
increases with the galaxy mass: in the last merging of equal mass
systems and of the accretion scenario, the total number of particles
involved is of the order of $5.2 \times 10^5$ and $3.2 \times 10^5$,
respectively.

%_____________________________________________________________
%                 A figure as large as the width of the column
%-------------------------------------------------------------
   \begin{figure} \centering
   \includegraphics[width=0.5\textwidth]{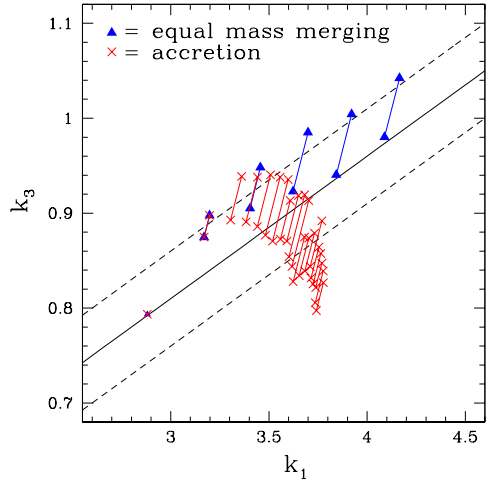}
\caption{The merging end--products in the ($\ku,\kt$) plane, where the
solid line represents the edge--on FP relation with its observed
1-$\sigma$ dispersion (dashed lines). Equal mass mergers are shown as
triangles; crosses represent the accretion hierarchy. Bars show the
amount of projection effects.} \label{FigVibStab}
\end{figure}
%
%_____________________________________________________________

%_____________________________________________________________
%                 A figure as large as the width of the column
%-------------------------------------------------------------
   \begin{figure}
   \centering
   \includegraphics[width=0.5\textwidth]{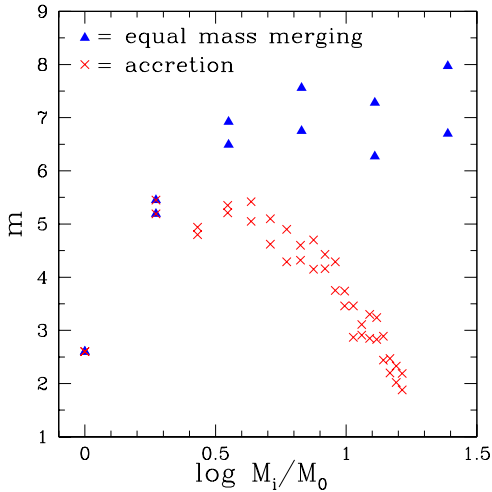}
\caption{Sersic best fit parameter $m$ vs. total stellar mass of the
end--products at stage $i$ of the merging hierarchy. Symbols are the
same as in Fig. 4} \label{FigVibStab} \end{figure}
%
%_____________________________________________________________

In Fig. 4 we plot the results of equal mass merging and accretion
simulations in the plane $(\ku,\kt)$, where the {\it observed} FP is
seen almost edge--on, with best--fit $\kt=0.15\ku+0.36$ (solid line)
and scatter ${\rm rms}(\kt)\simeq 0.05$ (dashed lines; see Bender,
Burstein \& Faber 1992). In the diagram the end--products of equal
mass mergers and accretion are identified by triangles and crosses,
respectively.  The (common) progenitor of the merging hierarchies
(point without bar) is placed on the edge--on FP.  The straight lines
in Fig. 4, associated with the merger remnants, represent the range
spanned in the ($\ku,\kt$) space by each end--product, when observed
over the solid angle, as a consequence of projection effects. It is
interesting to note that for all the end--products the projection
effects are of the same order as the observed FP dispersion.  In
addition it is also apparent from Fig. 4 how, in case of equal mass
merging, $\kt$ and $\ku$ increase with merging in a way full
consistent with the observed FP tilt and thickness.  On the other
hand, in the case of accretions, after few steps the end--products are
characterized by a $\kt$ {\it decreasing} for increasing $\ku$, at
variance with the FP slope and the trend shown by the end--products of
equal mass mergers.  As a consequence, the last explored models
(corresponding to an effective mass increase of a factor $\sim 16$)
are found well outside the FP scatter.

The different behavior of the end--products in the two scenarios is
due mainly to effects of {\it structural non--homology}. This is
clearly shown by an analysis of the projected stellar mass density
profiles of the end--products.  In fact, we fitted (over the radial
range $0.1 \, \lsim \, R/\cRe \, \lsim \, 4$) the projected density
profiles of the end--products with the Sersic (1968) $R^{1/m}$ law,
and in Fig. 5 we plot the best fit parameter $m$ as a function of the
total mass of the systems, for equal mass mergers (triangles) and
accretion (crosses). Clearly, the best-fitting $m$ depends on the
relative orientation of the line--of--sight and of the end--products
of the simulations: the two points for each value of the mass in
Fig. 5 show the range of values spanned by $m$ when projecting the
final states along the shortest and longest axis of their inertia
ellipsoids.  In case of equal mass merging, the trend is $m$
increasing with galaxy mass, as observed in real elliptical galaxies
(see, e.g., Bertin, Ciotti \& Del Principe 2002 and references
therein). On the contrary, in case of accretion the best Sersic $m$
parameter decreases at increasing mass of the end--product, a behavior
opposite to what is empirically found.

As it is well known, elliptical galaxies follow, besides the FP, also
other scaling relations, such as the Faber--Jackson (hereafter FJ,
Faber \& Jackson 1976), and the Kormendy (1977) relations. It is
remarkable that, in our simulations, both the equal mass merging and
the accretion hierarchies fail at reproducing these two relations. In
fact, the end--products have central velocity dispersion lower, and
effective radius larger than what predicted by the FJ and the Kormendy
relations, respectively. In case of equal mass mergers, these two
effects curiously compensate, thus preserving the edge--on FP. In
other words, the remnants of {\it dissipationless} merging like those
described, are {\it not} similar to real elliptical galaxies, though
in some cases they reproduce the edge--on FP (see NLC02 for details).

\section{Conclusions}

The main results of our work can be summarized as follows:

\begin{itemize}

\item Our implementation of the D02 scheme in a F-90 N-body code
(FVFPS) has effective $O(N)$ complexity for number of particles $N
\gsim 10^4$, though its time efficiency is lower by a factor of $\sim
1.5$ than Dehnen's C++ implementation.

\item As shown by numerical tests, the parallelized version of the
FVFPS code has good scaling with the number of processors (for example
for numbers of particles $N \simeq 10^6$, at least up to 16
processors).
 
\item We found very good agreement between the results of merging
simulations performed with our FVFPS code and with the GADGET code.

\item From an astrophysical point of view, our high resolution
simulations of hierarchies of {\it dissipationless} galaxy mergers
indicate that {\it equal mass} merging is compatible with the
existence of the FP relation. On the other hand, when an {\it
accretion} scenario is considered, the merger remnants deviate
significantly from the observed edge--on FP (see also NLC02).

\item The different behavior with respect to the FP of the merger
remnants in the two scenarios is a consequence of structural
non--homology, as shown by the analysis of their projected stellar
density profiles.

\item In any case, the end--products of our simulations fail to
reproduce both the FJ and Kormendy relations. In the case of equal
mass merging, the combination of large effective radii and low central
velocity dispersions maintains the remnants near the edge--on FP.

\end{itemize}

\begin{acknowledgements}
We would like to thank Walter Dehnen for helpful discussions and
assistance.  Computations have been performed at the CINECA center
(Bologna) under the INAF financial support. L.C. was supported by
MURST CoFin2000.
\end{acknowledgements}

\bibliographystyle{aa}

\end{document}